\documentclass[twocolumn,amssymb,amsmath,superscriptaddress,aps,pre,reprint]{revtex4-1}
\usepackage{graphicx}
\usepackage{dcolumn}
\usepackage{bm}
\usepackage{subfigure}

\begin{document}

\title{Mean field approximation of two coupled populations of excitable units}

\author{Igor Franovi\' c}
\affiliation{Faculty of Physics, University of Belgrade, PO Box 44, 11001 Belgrade, Serbia}

\author{Kristina Todorovi\' c}
\affiliation{Department of Physics and Mathematics,Faculty of Pharmacy, University of Belgrade,
Vojvode Stepe 450, Belgrade, Serbia}

\author{Neboj\v sa Vasovi\' c}
\affiliation{Department of Applied Mathematics, Faculty of Mining and Geology, University of Belgrade, PO Box 162, Belgrade, Serbia}

\author{Nikola Buri\' c}
\email{buric@ipb.ac.rs}
\affiliation{Scientific Computing Lab., Institute of Physics, University of Beograd, PO Box 68, 11080 Beograd-Zemun, Serbia}%

\date{\today}% It is always \today, today,

\begin{abstract}
The analysis on stability and bifurcations in the macroscopic dynamics exhibited by the system of two coupled large populations comprised of $N$ stochastic excitable units each is performed by studying an approximate system, obtained by replacing each population with the corresponding mean-field model. In the exact system, one has the units within an ensemble communicating via the time-delayed linear couplings, whereas the inter-ensemble terms involve the nonlinear time-delayed interaction mediated by the appropriate global variables. The aim is to demonstrate that the bifurcations affecting the stability of the stationary state of the original system, governed by a set of $4N$ stochastic delay-differential equations for the microscopic dynamics, can accurately be reproduced by a flow containing just four deterministic delay-differential equations which describe the evolution of the mean-field based variables. In particular, the considered issues include determining the parameter domains where the stationary state is stable, the scenarios for the onset and the time-delay induced suppression of the collective mode, as well as the parameter domains admitting bistability between the equilibrium and the oscillatory state. We show how analytically tractable bifurcations occurring in the approximate model can be used to identify the characteristic mechanisms by which the stationary state is destabilized under different system configurations, like those with symmetrical or asymmetrical inter-population couplings.
\end{abstract}

\pacs{02.30Ks, 05.45.Xt}

\maketitle

The onset and mutual adjustment of collective rhythms are regarded as the dynamical paradigm for the macroscopic phenomena in a wide range of biological and inorganic systems. Such a framework has already proven indispensable for understanding the normal and pathological patterns of brain activity \cite{neural,T07}, coordination of cellular clocks governing the circadian rhythms \cite{YIMOYK03}, the mechanisms regulating heartbeat \cite{GM88} or lying behind certain forms of social behavior \cite{social}, entrainment of electrochemical oscillators \cite{KZH02}, as well as the dynamics of Josephson junction circuits \cite{WCS96} and the arrays of coupled lasers \cite{KRACS05}. The emergence of macroscopic rhythms in ensembles of oscillating units is mediated by the synchronization based self-organization \cite{synchro}. The latter is often influenced or facilitated by noise on one hand \cite{noise,ampdeath}, while on the other hand, the interaction over the appropriate communication channels is typically susceptible to transmission delays or there may be a time lag due to the system components' latency in response to input variations \cite{delay}. A pervasive idea in nonlinear dynamics is to treat an assembly exhibiting a collective mode as a macroscopic oscillator \cite{BRZKP09}, which could in turn be subjected to an external drive or be exposed to a single or multiple collective rhythms from other populations. In this context, an important issue is to consider the relationship between the ensembles's global variable and the external forcing or that between the corresponding global variables.

In terms of the dynamical complexity of the observed behavior and the methods available for the analytical study, one has to make a distinction between the cases where the populations are built of self-sustained (autonomous) oscillators or the excitable units. In the former instance, it is possible to obtain a more compact description of the interacting ensembles' dynamics by applying the phase reduction techniques \cite{OA08,Kawamura,KNAKK09}. Given that the phase cannot be attributed to the system residing at the equilibrium, excitable populations are not amenable to such methods. Nonetheless, on the level of elementary behavior associated with the macroscopic variables, populations containing the excitable or self-oscillating units undergo qualitatively similar forms of dynamics. In particular, the ensuing collective modes may synchronize \cite{BHOS08,MKB04}, become phase-locked or get suppressed by the action of the coupling delay (delay-induced amplitude death) \cite{RP04}. Beyond such simple cases, there are more complex forms of collective behavior tied exclusively to populations of interacting oscillators. A few prominent examples include the self-organized quasiperiodicity \cite{BRZKP09} and the partially synchronous chimaera states \cite{AMSW08,OPT10}, which have been found to emerge in systems of identical phase oscillators under the action of external forcing or by coupling to another population, respectively. The former regime is characterized by the frequency of the collective mode being distinct from that of the single elements, while the other involves a broken symmetry between the dynamics of two interacting populations.

In this study, the focus lies with the two delay-coupled populations of identical excitable units modeled
by the Fitzhugh-Nagumo elements. The behavior of the latter is representative for the type II excitability \cite{I07}, which in contrast to type I, lacks a sharp threshold in a sense that the amplitude of the response depends continuously on the size of the applied stimulus. Though the considered framework is quite general, the basic motivation admittedly draws from the observations on neuronal assemblies, with the adopted model of local activity typically invoked in such a context. The analysis of the underlying system dynamics may be approached from two different angles. For one, a numerical study can be carried out to look for the states of the increasing dynamical complexity. Instead, we take on a strategy that consists in examining how well is the behavior of the exact system matched by that of the coupled mean-field (MF) systems, having derived the MF model as an approximation for the activity of a single ensemble. The concept aims to fully exploit the analogy between the assemblies and the macroscopic oscillators, such that the original set of equations for the microscopic dynamics is reduced to a flow which describes the evolution of the global variables, incorporating the cross-population interaction in a natural way. An important ingredient for the setup is that both the intra- and the cross-population coupling terms include the transmission delays. Note that the layout with two populations may constitute a paradigm, or rather serve as a nucleus for the "network of networks" \cite{BHOS08,OPT10,SR12}, which can be realized as a hierarchy of multiple networks, or it could be thought of as an idealization for a single network with a strong modular structure and a large number of elements in each community (subnetwork). Both configurations are common in biological systems \cite{BHOS08}, ranging from the cellular level to the distributed anatomical areas of the brain, and also encompassing the populations of cells responsible for the rhythmic activity in heart, kidney, pancreas, to name but a few. As for the comparison with the MF model, the attempts at providing a reduced description instead of using the complete set of equations for each and every population constituent, have a particularly long history within the neuroscience \cite{LGNS04,SJ08,H08}. Apart for the gains on the modeling side, they have initially been instigated by the finding that the EEG and MEG recordings may be linked to an average behavior, viz. the massively summed action potentials emitted within the strongly coupled, but remote cortical areas \cite{DF03,SJ08}. Though the given approach inevitably includes simplifying assumptions that eventually constrain the repertoire of possible system behaviors just to periodic motion, some of the realism may readily be sacrificed for a more parsimonious representation if the emergence of the collective mode and the related dynamics are reproduced with sufficient fidelity.

The mean field approximation has been applied on
systems of excitable units with noise but with no time-delay for
example in \cite{Takvel},\cite{Tanabe},\cite{Chaos},\cite{ZNFSG03}.
Otherwise a type of MF approximarion was devised in \cite{Hasagawa1} and
\cite{Hasagawa2} and applied on large clusters of noisy neurons
with time-delayed interaction in \cite{Hasagawa3}. However, the
approximations made in these papers resulted in a system of
equations that is still to large to be analyzed analytically, so
that the approximate system must be studied numerically. We shall
utilize an approximate system of only two DDDE, introduced in \cite{BRTV10}, for the dynamics of
the mean fields for each of the two populations.  Such a simple system allows analytical treatment
of bifurcations and the parameter domains of stability of the
stationary states which turn out to be in a quite good agrement
with the exact complex system.

 The key set of issues addressed in this study amounts to identifying the conditions for the stability of the stationary state, the onset of the collective mode, bistability between the equilibrium and the oscillation state, as well as the time-delay induced suppression of the collective mode. One notes that the applied term "collective mode" here implies the existence of a limit cycle for the total system of interacting populations. Though the intention is not, or rather cannot be to account for any experimental observation of such phenomena, some elementary comparison can still be drawn. For instance, the notion that the emergence and the synchronization properties of collective rhythms arising in the macroscopic neural populations are critically influenced by the coupling strength and the interaction delay \cite{DF03} has its clear analogue in the results we arrive at. Consistent with the stated objectives, the study of the approximate system is concerned with the local bifurcation analysis, carried out analytically and corroborated by the numerical means, to determine $i)$ the parameter domains of stability of the steady states, $ii)$ the scenarios for the emergence or the suppression of the collective mode, and $iii)$ the parameter domains admitting the bistability between the equilibrium and the oscillatory state.

The paper is organized as follows. In Section \ref{Model}, the details of the exact model of interacting
populations are laid out in parallel with the derivation of its MF counterpart. Section \ref{Results:analytical}
is focused on the local bifurcation analysis of the approximate model, providing for the analytical results. In Section \ref{Results:comparison}, we demonstrate that the approximation based on two coupled MF systems is able to accurately predict the behavior of the exact system in terms of the stability of the equilibrium, as well as the onset and the suppression of the collective mode. It is also pointed out how different system configurations affect the scenarios for the emergence of the oscillatory state and influence the parameter domains supporting its coexistence with the equilibrium. The results are briefly summarized and discussed in the concluding section.

\section{Background on the exact model and derivation of its MF counterpart}\label{Model}

\subsection{Details of the exact model}\label{sub:exact}

Each population comprises a collection of $N$ identical Fitzhugh-Nagumo elements \cite{I07,diffusive}, whose dynamics is given by
\begin{align}
\epsilon dx_{i,1}&=(x_{i,1}-x_{i,1}^3/3-y_{i,1}+I_1)dt+\frac{g_{in,1}}{N}\times\nonumber \\
&\sum_{j=1}^{N}[x_{j,1}(t-\tau_{in,1})-x_{i,1}(t)]dt+g_{c,1}\times\nonumber \\
&\arctan[X_2(t-\tau_{c,1})+b_2]dt, \nonumber\\
dy_{i,1}&=(x_{i,1}+b_1)dt + \sqrt{2D_1}dW_{i,1}\nonumber\\
\epsilon dx_{i,2}&=(x_{i,2}-x_{i,2}^3/3-y_{i,2}+I_2)dt+\frac{g_{in,2}}{N}\times\nonumber \\
&\sum_{j=1}^{N}[x_{j,2}(t-\tau_{in,2})-x_{i,2}(t)]dt+g_{c,2}\times\nonumber \\
&\arctan[X_1(t-\tau_{c,2})+b_1]dt, \nonumber\\
dy_{i,2}&=(x_{i,2}+b_2)dt + \sqrt{2D_2}dW_{i,2}, \label{eq1}
\end{align}
where the subscripts $k=1,2$ specify the population, indices $i=1,..N$ denote a particular
unit within the population, and $X_k=(1/N)\sum\limits_{i=1}^{N}x_{i,k}$ stand for the macroscopic
variables that typify the global population behavior. The small parameter $\epsilon=0.01$ imposes
a wide separation between the characteristic time scales for the evolution of $x_{i,k}$ and $y_{i,k}$.
In the context of neuronal activity the set of fast variables embodies the membrane potentials, whereas the slow-variable set is supposed to account for the gross kinetics of the potassium ion-gating channels. In the
absence of an external stimulation $I_1=I_2=0$ applies. The impact of a noisy background activity is reflected by the $\sqrt{2D}dW_{i}$ terms, which represent the stochastic increments of the independent Wiener processes specified by the noise amplitude $D$, expectation values $\langle dW_i \rangle=0$ and the correlations that satisfy $\langle dW_idW_j\rangle=\delta_{ij}dt$ for each population.

Owing to the system configuration, the local dynamics involves two types of interactions, each characterized by the coupling strength and the delay. The respective parameters associated with the intra-ensemble terms are $g_{in,k}$ and $\tau_{in,k}$, while the cross-population terms are awarded $g_{c,k}$ and $\tau_{c,k}$. Within the populations, the elements communicate via the simple linear (diffusive) couplings, such that $\tau_{in}$ may account for the transmission delays due to finite rate of signal propagation or the latency in unit responses. Given the objectives stated in the Introduction, it is not unjustified to make use of some simplifying assumptions, like the all-to-all pattern of interconnections and the uniformity of coupling strengths inside the ensembles, which are the abstractions often invoked in the relevant literature \cite{alltoall}. As for the cross-population interactions, at the current stage no particular model is considered to be preferred over the others. However, we make use of an analogy to neural systems by noting how a variety of models display a common feature. Stated in the language of neuroscience, the evoked postsynaptic potentials can be expressed in a symbolical form $h=s\otimes m$, where $m$ refers to an average density of presynaptic input arriving from the transmitter population, and $s$ presents the threshold-like response of the neurons of the receiving population \cite{DF03}. Adhering to this concept, the output of the transmitter population is integrated by the macroscopic variables $X_k=(1/N)\sum\limits_{i=1}^{N}x_{i,k}$, which reflect the global behavior in a sense that the better the synchronization among the constituent elements, the larger the amplitudes of $X_k$. In terms of the nonlinear threshold function, there is a degree of arbitrariness, so the $\arctan$ form applied here is as good a choice as any. Unlike the interactions within the populations, which are characterized by the specific strengths per link, the inter-population terms involve the cumulative strengths, consistent with the idea of viewing each population as a single macroscopic oscillator. The meaning of the parameter $b$ is explained in more detail further below. The bidirectional couplings between the ensembles, being either symmetrical or asymmetrical, may be important from the aspect of neuroscience, given that the brain connectivity patterns are known to exhibit a large portion of reciprocal interactions \cite{SJ08}. On the level of local dynamics, the parameter $b$ plays a key role as it modulates the unit's excitability. For an isolated unit in the noiseless case, the condition $|b|=1$ determines the Hopf bifurcation threshold, above which the system possesses a unique equilibrium, whereas below it one finds a limit cycle. Selecting $b$ slightly above $1$, like the value $b=1.05$ held throughout the paper, the population elements are poised quite close to the Hopf threshold, which gives rise to an excitable behavior. In such a regime, an adequate stimulation, be it by the noise or the interaction terms, may evoke large transients within the fast variable subspace before the ensuing trajectories converge back to rest. Note that in the scenario where noise acts in the slow subsystem, the elicited limit cycles are just the precursors of the deterministic ones \cite{FHNdyn}. Turning back to the role of $b$ in the cross-population coupling, it is seen to ensure that the largest contribution to the interaction term comes from the global states lying farthest away from the equilibrium.

\subsection{Background and the formulation of the MF approximation} \label{sub:MF}

Deriving the MF approximation, we aspire for a highly reduced set of nonlinear DDE instead of the original system
\eqref{eq1} comprised of a large set of nonlinear SDDE. Though a simplified representation, the MF model should still be able to reproduce with sufficient accuracy the latter's behavior regarding the stability of the steady states, the scenarios for the onset of the collective mode and its suppression under the action of the cross-population coupling delay. The MF treatment draws on the all-to-all type of connectivity among neurons within each population, incorporating the thermodynamic limit $N\rightarrow\infty$ in a natural way \cite{LGNS04}. In order to build a MF model, two different approaches are available to proceed with: one may either consider the time-dependence of a hierarchy of probability densities according to the Fokker-Planck formalism, or may focus on the evolution of cumulants, whereby the full density of states is factorized into a series of marginal densities. The latter alternative is preferred, as it allows for a number of convenient approximations to be introduced in a controlled fashion \cite{LGNS04}. Note how one is bound to make some approximations for the nonlinearity of the original system, given that the cumulants of the particular order are usually linked to those of the higher order, which apparently renders the underlying series unclosed. The way to resolve this issue consists in truncating the series by a form of a closure hypothesis. Such hypothesis typically integrates the cumulant approach with the Gaussian approximation \cite{Gaussian1,Gaussian2}, recalling that the Gaussian distribution has vanishing cumulants above the second order.

Confined to a single population, the Gaussian approximation involves two elementary prepositions: first, that the instantaneous distributions of local variables $P(x_i)$ and $P(y_i)$ are Gaussian, and second, that the ensemble
averages at any given moment coincide with the expectation values of the appropriate distributions in a sense
$(1/N)\sum\limits_{i=1}^{N}x_i\approx E[P(x_i)], (1/N)\sum\limits_{i=1}^{N}y_i\approx E[P(y_i)]$ \cite{Gaussian1,Gaussian2}. If the two stated conditions are met, all the cumulants above the second order are supposed to vanish. Let us briefly comment on the constraints which these conditions impose on the system parameters. As for the first point, the Gaussian distribution of local variables is maintained if the noise amplitude obeys $D<<1$. Nonetheless, the strong law of large numbers \cite{A98} implies that the second condition concerning the ensemble averages is fulfilled exactly in the thermodynamic limit $N\rightarrow\infty$ if the involved stochastic processes are independent $(g_{in}<<1)$. However, the numerical results presented further on indicate that the MF approximation remains valid if the two latter conditions are relaxed, viz. when there is non-negligible interaction in the finite-size systems, provided that the requirement for not too large a noise amplitude is satisfied.

In the following, we outline the key steps in the derivation of the MF model for the activity of an interacting assembly.
 The derivation presents a slight generalization of the one presented in \cite{BRTV10}.  To begin with, note that the cross-population coupling terms involve only the average dynamics of the respective transmitter populations. This means that the focus should really lie with the internal ensembles' dynamics, treating them temporarily as if they were independent, while subsequently including the inter-population
interaction. Therefore, we confine further presentation to a single population, whose dynamics is extracted from
\eqref{eq1} by setting $g_{c,1}$ or $g_{c,2}$ to zero
\begin{align}
\epsilon dx_i&=(x_i-x_i^3/3-y_i)dt+\frac{g_{in}}{N}\sum_{j=1}^{N}[x_j(t-\tau_{in})-x_i(t)]dt, \nonumber\\
dy_i&=(x_i+b)dt + \sqrt{2D}dW_i, \label{eq2}
\end{align}
Given that the distributions of the stochastic local variables are assumed to take on the Gaussian form, one can fully characterize them by the set of the first and second order moments, which includes the mean values, the variances and the covariance. The mean values applied here
\begin{align}
m_x(t)&=\langle x_i(t)\rangle=\lim\limits_{N\rightarrow\infty}(1/N)\sum\limits_{i=1}^{N}x_i(t) \nonumber \\ m_y(t)&=\langle y_i(t)\rangle=\lim\limits_{N\rightarrow\infty}(1/N)\sum\limits_{i=1}^{N}y_i(t) \label{eq3}
\end{align}
should strictly speaking be distinguished from the global variables $X$ and $Y$ considered earlier for the large, but still finite-size populations. The angled brackets are generally used to denote averaging over the units making up the ensemble, whereas $m_x$ and $m_y$ are reserved solely for the averages of the local variables. Before introducing the second order moments, it is convenient to define the deviations from the mean
$n_{x_i}(t)=\langle x_i(t)\rangle-x_i(t)$ and $n_{y_i}(t)=\langle y_i(t)\rangle-y_i(t)$, which obey the Gaussian distributions and are independent between the single elements. Then the appropriate variances read
\begin{align}
s_x(t)&=\langle n_{x_i}^2(t)\rangle=\langle(\langle x_i(t)\rangle-x_i(t))^2\rangle \nonumber \\
s_y(t)&=\langle n_{y_i}^2(t)\rangle=\langle(\langle y_i(t)\rangle-y_i(t))^2\rangle, \label{eq4}
\end{align}
whereas the covariance is given by
\begin{equation}
u(t)=\langle n_{x_i}(t)n_{y_i}(t)\rangle=\langle(\langle x_i(t)\rangle-x_i(t))(\langle y_i(t)\rangle-y_i(t))\rangle. \label{eq5}
\end{equation}

The evolution of the distributions' means $m_x$ and $m_y$ is obtained by performing the ensemble averages
over the system (\ref{eq2}), while the expressions for the dynamics of $s_x,s_y$ and $u$ follow from explicitly
taking the time derivatives of the definitions (\ref{eq4}) and (\ref{eq5}). Note that the latter calculation also involves the derivatives of the compound functions of the stochastic variables such as $d\langle x_i^2\rangle/dt$ and $d\langle y_i^2\rangle/dt$, where one is required to apply the Ito's chain rule. As for the higher order averages, like $\langle x_i^2\rangle$ and $\langle x_i^3\rangle$, it is necessary to tie them to the first and second order
moments. In the simplest cases, this is accomplished by using the definitions (\ref{eq4}) and (\ref{eq5}), while in most instances one arrives at the required relations by setting the higher order cumulants \cite{G85} to zero, e.g.
$\langle x_i^3\rangle_c=\langle x_i^3\rangle-3\langle x_i^2\rangle\langle x_i\rangle+2\langle x_i\rangle^3=0$,
$\langle x_i^2y_i\rangle_c=\langle x_i^2y_i\rangle-\langle x_i^2\rangle\langle y_i\rangle-2\langle x_i\rangle\langle x_iy_i\rangle+2\langle x_i\rangle^2\langle y_i\rangle=0$, and similar for $\langle x_i^3y_i\rangle_c=0$ and
$\langle x_i^4\rangle_c=0$. The ensuing auxiliary formulas for the higher order averages then read
\begin{align}
\langle x_i^2\rangle&=s_x+m_x^2 \nonumber \\
\langle x_i^3\rangle&=m_x^3+3m_xs_x \nonumber \\
\langle x_i^4\rangle&=m_x^4+6m_x^2s_x+3s_x^2 \nonumber \\
\langle x_iy_i\rangle&=u+m_xm_y \nonumber \\
\langle x_i^2y_i\rangle&=m_ys_x+m_ym_x^2+2m_xu \nonumber \\
\langle x_i^3y_i\rangle&=3s_xu+3m_x^2u+m_ym_x^3+3m_xm_ys_x. \nonumber \\ \label{eq6}
\end{align}
After a series of steps which are too lengthy to convey in full detail, the closed system of equations for the first and second order moments finally becomes
\begin{align}
\epsilon \frac{dm_x(t)}{dt}&=m_x(t)-m_x(t)^3/3-s_x(t)m_x(t)-m_y(t)+\nonumber \\
&g_{in}(m_x(t-\tau_{in})-m_x(t)) \nonumber \\
\frac{dm_y(t)}{dt}&=m_x(t)+b \nonumber \\
\frac{\epsilon}{2}\frac{ds_x(t)}{dt}&=s_x(t)(1-m_x^2(t)-s_x(t)-g_{in})-u(t) \nonumber \\
\frac{1}{2}\frac{ds_y(t)}{dt}&=u(t)+D \nonumber \\
\frac{du(t)}{dt}=&\frac{u(t)}{\epsilon}(1-m_x^2(t)-s_x(t)-g_{in})-\frac{1}{\epsilon}s_y(t)+s_x(t). \label{eq7}
\end{align}
Note that (\ref{eq7}) comprises a set of deterministic delay equations, where the impact of noise is absorbed into
its amplitude $D$. Recalling the Introduction, one of the objectives has been to carry out the bifurcation analysis on the MF model analytically. However, the system (\ref{eq7}) is still sufficiently complex to defy such a treatment. To ensure that the bifurcation analysis is analytically tractable, we consider an additional approximation which concerns the relatively fast relaxation of the second order moments. Given that the characteristic time scales, at least for $s_x$ and $u$, are dominated by the small parameter $\epsilon$, one may substitute their full dynamics by the stationary values reached when $\dot{s_x}=0,\dot{s_y}=0$ and $\dot{u}=0$ are satisfied. Though a crude approximation, it is not an uncommon one \cite{LGNS04,Gaussian2}. In the language of neuroscience, the net result it yields is comparable to translating the initial MF model into an effective neural-mass model \cite{DF03}, the former (latter) associated with the system of five (two) equations. Nevertheless, whether this is justified or not strongly depends on the main objectives of the study, which here concern the stability of the stationary state, the onset of the collective mode and its suppression in an amplitude death-like phenomenon \cite{ampdeath}. As it stands, the described modification to the MF model should not substantially affect the latter set of issues, since the information supplied by the second order variables, like that on small fluctuations around the collective synchronous state, appears redundant in such a context. This is corroborated later on by the results indicating an agreement between the behaviors of the exact and the MF approximation.

To complete the MF approximation for the dynamics of the two interacting populations, one should take into account the inter-ensemble interactions initially left aside, arriving at the following set of four equations
\begin{align}
\epsilon \frac{dm_{x,1}(t)}{dt}&=m_{x,1}(t)-\frac{m_{x,1}(t)^3}{3}-\frac{m_{x,1}(t)}{2}(1-g_{in,1}- \nonumber \\
&m_{x,1}(t)^2+\sqrt{(g_{in,1}-1+m_{x,1}(t)^2)^2+4D_1})- \nonumber \\
&m_{y,1}(t)+g_{in,1}(m_{x,1}(t-\tau_{in,1})-m_{x,1}(t))+ \nonumber \\
&g_{c,1}\arctan(m_{x,2}(t-\tau_{c,1})+b_2) \nonumber \\
\frac{dm_{y,1}(t)}{dt}&=m_{x,1}(t)+b_1 \nonumber \\
\epsilon \frac{dm_{x,2}(t)}{dt}&=m_{x,2}(t)-\frac{m_{x,2}(t)^3}{3}-\frac{m_{x,1}(t)}{2}(1-g_{in,2}- \nonumber \\
&m_{x,2}(t)^2+\sqrt{(g_{in,2}-1+m_{x,2}(t)^2)^2+4D_2})- \nonumber \\
&m_{y,2}(t)+g_{in,2}(m_{x,2}(t-\tau_{in,2})-m_{x,2}(t))+ \nonumber \\
&g_{c,2}\arctan(m_{x,1}(t-\tau_{c,2})+b_1) \nonumber \\
\frac{dm_{y,2}(t)}{dt}&=m_{x,2}(t)+b_2 \nonumber \\ \label{eq8}
\end{align}
Note that for $D_1=D_2=0$, the obtained system strongly resembles the case of two interacting Fitzhugh-Nagumo
elements subjected to the delayed feedback. Another point is that the isolated populations ($g_{c,1}=g_{c,2}=0$)
can be shown to exhibit the excitable-like dynamics under the variation of $D$ and $\tau$. By this is meant that apart from the small amplitude oscillations about the equilibrium, there may also be large excursions of the global potential, this reflecting the crucial feature of the exact system. In our previous paper, it has already been demonstrated that the MF model of a single assembly is able to accurately predict the qualitative behavior of the exact system \cite{BRTV10}. This refers to a sequence of local bifurcations under variation of $D$, $\tau_{in}$ and $g_{in}$, which can be used to highlight the parameter domains giving rise to oscillatory states or those that lead to the amplitude death \cite{BTV10}. In addition, the MF model of an isolated ensemble has also been found to reflect the global bifurcation imminent to the onset of clustering in the exact system \cite{FTVB12}.

Before proceeding to the main results, several brief remarks on the applied numerical integration schemes are in order. The time series for both the exact and the approximate models are obtained by implementing the Euler method
with the fixed time step $\Delta t=0.005$ in the former, and $\Delta t=0.01$ in the latter case, having verified that no changes occur for the smaller $\Delta t$. Also, on either occasion, we have adopted the standard and physically plausible initial functions, based on the assumption of the units evolving independently within the time interval
$t\in[-\tau_{min},0]$, where $\tau_{min}=min\{\tau_{in,1},\tau_{in,2},\tau_{c,1},\tau_{c,2}\}$. This effectively amounts to integrating the systems (\ref{eq1}) and (\ref{eq8}) by disregarding any interaction for $t\in[-\tau_{min},0]$, with the initial conditions in each instance taken in the vicinity of the fixed point. The results for the exact model refer to populations made up of $N=200$ elements, but have been verified to persist if the larger assemblies are considered.

\section{Analytical results of the local bifurcation analysis of the approximate system} \label{Results:analytical}

In the two following sections, we first provide the details of the local bifurcation analysis performed on the  approximate system and then examine whether and how well do these results match the behavior of the exact system, whereby the latter dynamics is represented by the typical sample paths obtained from numerical integration of (\ref{eq1}) for the sufficiently large $N$ with $D_1,D_2\neq0$. On the first part, the analysis covers the stability of the attractor states for the total system of coupled populations, such that both of them are either found lying in the equilibrium or exhibiting oscillations. The main focus is on the stability of the fixed point and its destabilization under variation of the cross-population coupling strengths and delays. Apart for the onset of the oscillatory state, it is also considered how the coherent rhythms may become suppressed, this primarily attributed to the action of the inter-ensemble time lags. As a final matter, we demonstrate the existence of the parameter domains admitting the bistable regime, where the stationary and the oscillatory state coexist. Altogether, an inference confirmed later on is that the MF approximation can capture the behavior of the exact system much better if the collective dynamics is such that the deterministic component, controlled by the coupling strength and time delay, prevails over the stochastic component. The points enumerated above exhaust the corpus of problems that may approximately be treated by the local bifurcation theory, in a sense of explaining the qualitative changes arising in the system's asymptotic dynamics due to parameter variation. Outside the scope remain the more complex phenomena occurring for larger $D$-s and $\tau$-s, which could cause the behavior of single units within the populations to become substantially stratified. Such issues would fall under the notion of stochastic bifurcations \cite{A98}, meaning that one should consider how the parameter modification influences the changes of the respective stationary distributions of the local variables.

Since we discuss the scenarios with symmetrical and asymmetrical cross-population couplings, as well as the setups
where the inherent ensemble dynamics is the same or distinct, the analytical results of the local bifurcation analysis on the system of interacting MF models are presented in most general terms with respect to the system parameters. First, it is established that the system \eqref{eq8} possesses a unique equilibrium given by
\begin{align}
m_{x,k}=-b_k; m_{y,k}&=\frac{b_k}{2}[1+\frac{b_k^2}{3}+g_{in,k}- \nonumber \\
&\sqrt{(g_{in,k}-1+b_k^2)^2+4D_k}] \label{eq9}
\end{align}
with $k=1,2$. The local stability of \eqref{eq9} depends on the roots of the characteristic equation of the system \eqref{eq8}. To obtain the latter, one linearizes \eqref{eq8} around the equilibrium, assuming that the deviations are of the form $\delta m_{x,k}(t)=A_ke^{\lambda t}, \delta m_{y,k}(t)=B_ke^{\lambda t}$ and
$\delta m_{x,k}(t-\tau_{in,k})=A_ke^{\lambda (t-\tau_{in,k})}$. This results in a set of algebraic equations for the coefficients $A_k$ and $B_k$, which has a nontrivial solution only if
\begin{equation}
\Delta_1(\lambda)\Delta_2(\lambda)-\lambda^2g_{c,1}g_{c,2}e^{-\lambda(\tau_{c,1}+\tau_{c,2})}=0 \label{eq10}
\end{equation}
is fulfilled, where $\Delta_k(\lambda)=-\lambda F_k+\epsilon\lambda^2-g_{in,k}\lambda e^{-\lambda\tau_{in,k}}+1$
with $F_k=F_k(g_{in,k},b_k,D_k)$. The condition \eqref{eq10} poses the desired characteristic equation, whose being transcendental reflects the presence of (multiple) time delays in \eqref{eq8}. Though \eqref{eq10} has an infinite number of roots, it is well known how there may be only a finite number of exceptional roots equal to zero or with a zero real part \cite{HL93,Campbell}. One recalls that tangent to the subspace spanned by the associated eigenvectors lies the center manifold \cite{W00,K04}, where the qualitative features of the system's dynamics, such as the local stability, are contingent on the nonlinear terms.

Bifurcations of the stationary state take place for the parameter values where the roots of \eqref{eq10} cross the imaginary axes. Given that Eq. \eqref{eq10} does not admit the possibility $\lambda=0$, we look for the pure imaginary roots of the form $\lambda=\imath\omega$, adopting $\omega$ to be real and positive. Substituting for $\lambda$ in \eqref{eq10}, one obtains
\begin{align}
&[-\imath\omega(F_1-\imath\epsilon\omega+g_{in,1}(\cos\omega\tau_{in,1}-\imath\sin\omega\tau_{in,1}))+1]\times\nonumber\\
&[-\imath\omega(F_2-\imath\epsilon\omega+g_{in,2}(\cos\omega\tau_{in,2}-\imath\sin\omega\tau_{in,2}))+1]+\nonumber\\
&\omega^2g_{c,1}g_{c,2}(\cos(\omega(\tau_{c,1}+\tau_{c,2}))-\imath\sin(\omega(\tau_{c,1}+\tau_{c,2})))=0 \label{eq11}
\end{align}
which, after equating both the real and the imaginary parts with zero, provides for the implicit relations of
$\omega$ and the system parameters
\begin{align}
-\omega^2P_1P_2+Q_1Q_2&=-\omega^2g_{c,1}g_{c,2}\cos(\omega(\tau_{c,1}+\tau_{c,2})) \nonumber \\
\omega P_1Q_2+\omega P_2Q_1&=\omega^2g_{c,1}g_{c,2}\sin(\omega(\tau_{c,1}+\tau_{c,2})), \label{eq12}
\end{align}
where
\begin{align}
P_k&=F_k+g_{in,k}\cos(\omega\tau_{in,k}) \nonumber \\
Q_k&=\epsilon\omega^2+g_{in,k}\omega\sin(\omega\tau_{in,k})-1 \label{eq13}
\end{align}
applies for $k=1,2$. Squaring and adding the relations \eqref{eq12}, one arrives at
\begin{equation}
(\omega^2P_1P_2-Q_1Q_2)^2+\omega^2(P_1Q_2+P_2Q_1)^2=\omega^4g_{c,1}^2g_{c,2}^2, \label{eq14}
\end{equation}
which can be used to express the cross-population coupling strengths in terms of $\omega$, while keeping the
values for the subset of the intrinsic parameters $g_{in,k},\tau_{in,k},b_k$ and $D_k$ fixed. Obtained in a similar
fashion, the analogous result for the critical cross-population coupling delays may be written in the compact form
\begin{equation}
\tau_{c,1}+\tau_{c,2}=\frac{1}{\omega}\arctan(\frac{\omega P_1Q_2+\omega P_2Q_1}{\omega^2P_1P_2-Q_1Q_2}). \label{eq15}
\end{equation}
The last two equations combined define the curves in the appropriate delay-strength parameter plane. Bear in mind that Eq. \eqref{eq15} actually defines multiple branches of the Hopf bifurcation curves, these given by $\tau_{c,1}+\tau_{c,2}+j\pi$, where $j=0,1,2...$. Naturally, the above relations further simplify once the inter-ensemble couplings are taken to be symmetrical and/or the populations' intrinsic parameters are assumed to be identical. Note that the expressions such as these could not be obtained if we were to retain the initial MF model \eqref{eq7} containing the full dynamics of the second order cumulants. As for the type of bifurcations whose location is indicated by \eqref{eq15}, the very form of the solution adopted for the characteristic equation is consistent with the Hopf bifurcations, though a rigorous proof would require one to verify whether the conditions on non-hyperbolicity, transversality and genericity are satisfied \cite{K04,I07,Campbell}. Without entering into unnecessary details, it suffices to say that the system \eqref{eq8} admits both the supercritical and subcritical Hopf bifurcations, whereby the former (latter) result in the creation of a stable (unstable) limit cycle. In addition, recall that either of these types can be direct or inverse \cite{W00}, depending on whether an unstable two-dimensional manifold for the fixed point \eqref{eq8} appears or vanishes when crossing the bifurcation curve, respectively, having the fixed point unfold on the unstable or the stable side. The results derived analytically are corroborated numerically by means of the DDE-biftool \cite{biftool}, an adaptable package of Matlab routines suitable for handling the sets of DDE with constant delays.

\section{Qualitative comparison between the dynamics of the exact and the approximate system}\label{Results:comparison}

For the systematic study, we first consider the layout with two populations made up of independent excitable
elements ($g_{in,1}=g_{in,2}=0$) subjected to a common, comparably small noise ($D_1=D_2=0.0001$), whereby the cross-population coupling terms are taken to be symmetrical, so one may introduce $g_{c,1}=g_{c,2}=g_c$ and
$\tau_{c,1}=\tau_{c,2}=\tau_c$. The parameters are such that for $g_c=0$, the populations exhibit the asymptotically (stochastically) stable equilibrium in the MF (exact) model. Though it appears marginal at first sight, the described setup is still important, since the MF model is here strongly indicated to match the behavior of the real system. In a sense, this scenario is reminiscent of a null-hypothesis, given that the stated parameters are fully compliant with the nominal conditions for the validity of the MF approximation. One would further expect to gain some insight into the phenomena occurring for the more complex system configurations, or may at least obtain a reference point to isolate the effects of certain parameters, such as $g_{in}$ or $\tau_{in}$. In the remainder, the bifurcation diagrams are accompanied by the close-up views focused on the most relevant parameter domains, having those referred to in the text indicated by the representative symbols. Also, to distinguish between the different bifurcation curves, each is awarded with two types of indices. The $+/-$ sign specifies whether the curves coincide with the direct or inverse bifurcations, respectively, while the numerical index points to the order in which the given branches are encountered as the inter-population coupling delay is increased.

\begin{figure*}[t]%2
%\begin{minipage}[t]{\columnwidth}
\centering
\includegraphics[scale=0.4]{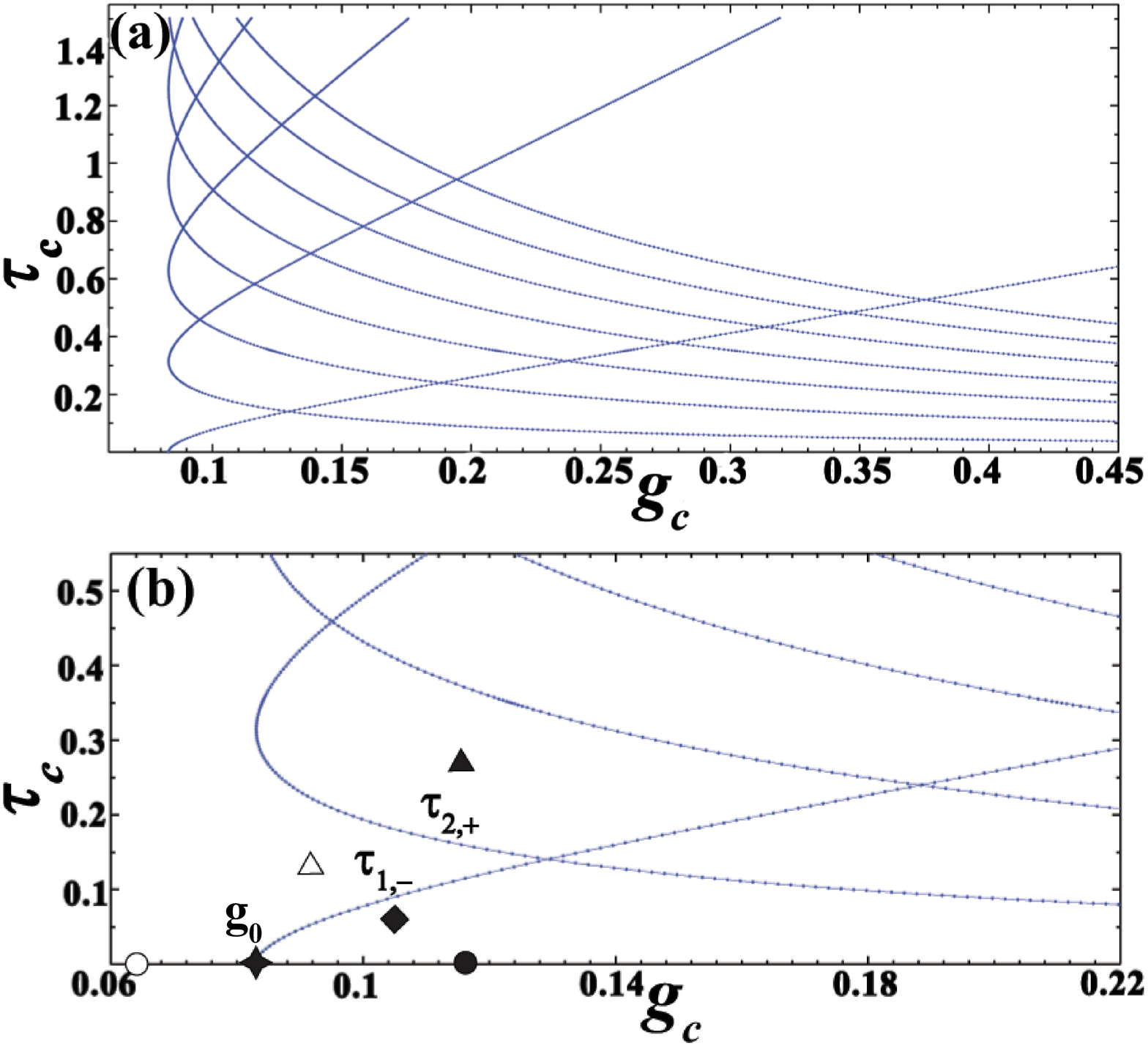}
%\end{minipage}
\caption{(color online) The first few branches of the Hopf bifurcation curves $\tau_c(g_c)$ are shown for the MF based approximation, derived for the system of two symmetrically coupled populations made up of independent excitable units ($g_{in}=0$). The latter part is a deliberate idealization, intended to probe the predictions of the approximate system when all the conditions required for the validity of the MF model are exactly satisfied. The onset of the collective mode is an instance of synchronization among elements receiving the common input. The units within the ensembles are subjected to noise whose amplitude is $D=0.0001$.
\label{Fig1}}
\end{figure*}

Appreciating the bifurcation diagram in Fig. \ref{Fig1}(a), a major point concerns the prediction on the existence of the critical strength $g_0$ for the instantaneous couplings ($\tau_c=0$), where the stationary state loses stability. For $g_c<g_0$, viz. the open circle in Fig. \ref{Fig1}(b), the equilibrium is stable, whereas for $g_c>g_0$ (solid circle) there is only the oscillatory state. The bifurcation scenario coincides with the direct supercritical Hopf bifurcation, and the numerical simulations imply that the unstable manifold for the equilibrium $m_{x,1}=m_{x,2}$ and $m_{y,1}=m_{y,2}$ around $g_c=g_0$ supports the oscillations in-phase, this being an example of synchronization between the units due to a common input. By the term "oscillations in-phase", it is meant that the MF approximation indicates a solution with the exact synchronization between the global variables, which is stochastically perturbed in the exact system. What is described applies not only for $\tau_c=0$, but also holds in any instance when the curve $\tau_{1,-}$ is crossed in the direction of increasing $g_c$ with $\tau_c$ kept fixed. However, we note that there is an additional subtlety to this transition derived from an interplay with the fold-cycle bifurcation, a global event which cannot be accounted for by the present type of analysis. One finds that the system undergoes a global bifurcation around $g_c\approx0.055$, by which an unstable and a large stable limit cycle are born. This witnesses of an interval $g_c<g_0$ where the stationary and the oscillatory state coexist, with their attraction basins separated by the unstable limit cycle. Above $g_0$, the incipient limit cycle emerges around the former position of the equilibrium, but cannot fully grow with supercriticality, given that it is enclosed by the unstable limit cycle created in the global bifurcation. At some point, the stable and the unstable limit cycle collide and disappear in an inverse fold-cycle bifurcation, which is indicative of a narrow $g_c$ interval around $\tau_{1,-}$ corresponding to a bistable regime with a small and a large amplitude limit cycle, viz. the solid diamond in Fig. \ref{Fig1}(b). However, such bistability may be difficult to observe in the exact system  for the sensitivity of the incipient cycle to stochastic perturbation, as even the very small noise amplitudes can prove sufficient to force the ensuing orbits outside of its neighborhood. Away from $g_0$, the system's trajectory is eventually drawn to a distant limit cycle attractor.

To proceed with, we consider the effects of increasing $\tau_c$ under fixed coupling strength $g_c>g_0$. Crossing the first bifurcation curve from below $\tau_c>\tau_{1,-}$, viz. the domain indicated by an open triangle in Fig. \ref{Fig1}(b), the equilibrium is seen to regain stability via the inverse supercritical Hopf bifurcation. Given the analogy of the underlying mechanisms, this could have been interpreted as a genuine example of the delay-induced amplitude death, if there were not for the large limit cycle which is unaffected by the local bifurcation. Instead, this stability domain is characterized by the coexistence between the stationary and the oscillatory state. Nonetheless, enhancing the delay above $\tau_{2,+}$ gives rise to a region of instability, represented by the solid triangle in Fig. \ref{Fig1}(b), where one encounters only the two populations oscillating in phase. Such an outcome is due to a supercritical Hopf bifurcation, which is reflected by the equilibrium gaining an unstable plane. Note that the analysis cannot extend to larger delays, since the underlying phenomena do not fall within the framework of the current study. It should be emphasized that the oscillation frequency of the MF model has been verified to match the one of the exact system almost perfectly. This point applies for two parameter domains highlighted by the solid circle and the solid triangle in Fig. \ref{Fig1}(b). Under $\tau_{1,-}$, the respective oscillation period of the approximate model is $T_{\bullet,MF}=3.664$ in arbitrary units, whereas the associated average period for the exact system is $T_{\bullet,EX}=3.668$. Likewise, in the domain instantiated by the solid triangle, $T_{\blacktriangle,MF}=3.874$ and $T_{\blacktriangle,EX}=3.869$. The cited data witness that the MF model is able to predict the average frequency of macroscopic oscillations of the exact system with remarkable accuracy. From the aspect of comparison between the real and the approximate systems, one should as well look back at the values of the critical strength $g_0$. The agreement here is weaker, whereby the MF model is found to overestimate the value. This is not unexpected, given that the local phenomena are mediated by the background global bifurcation. Still, the tendency and rate by which $g_0$ decreases with enhancing $D$ is reflected reasonably well by the MF model.

\begin{figure*}[t]%2
%\begin{minipage}[t]{\columnwidth}
\centering
\includegraphics[scale=0.4]{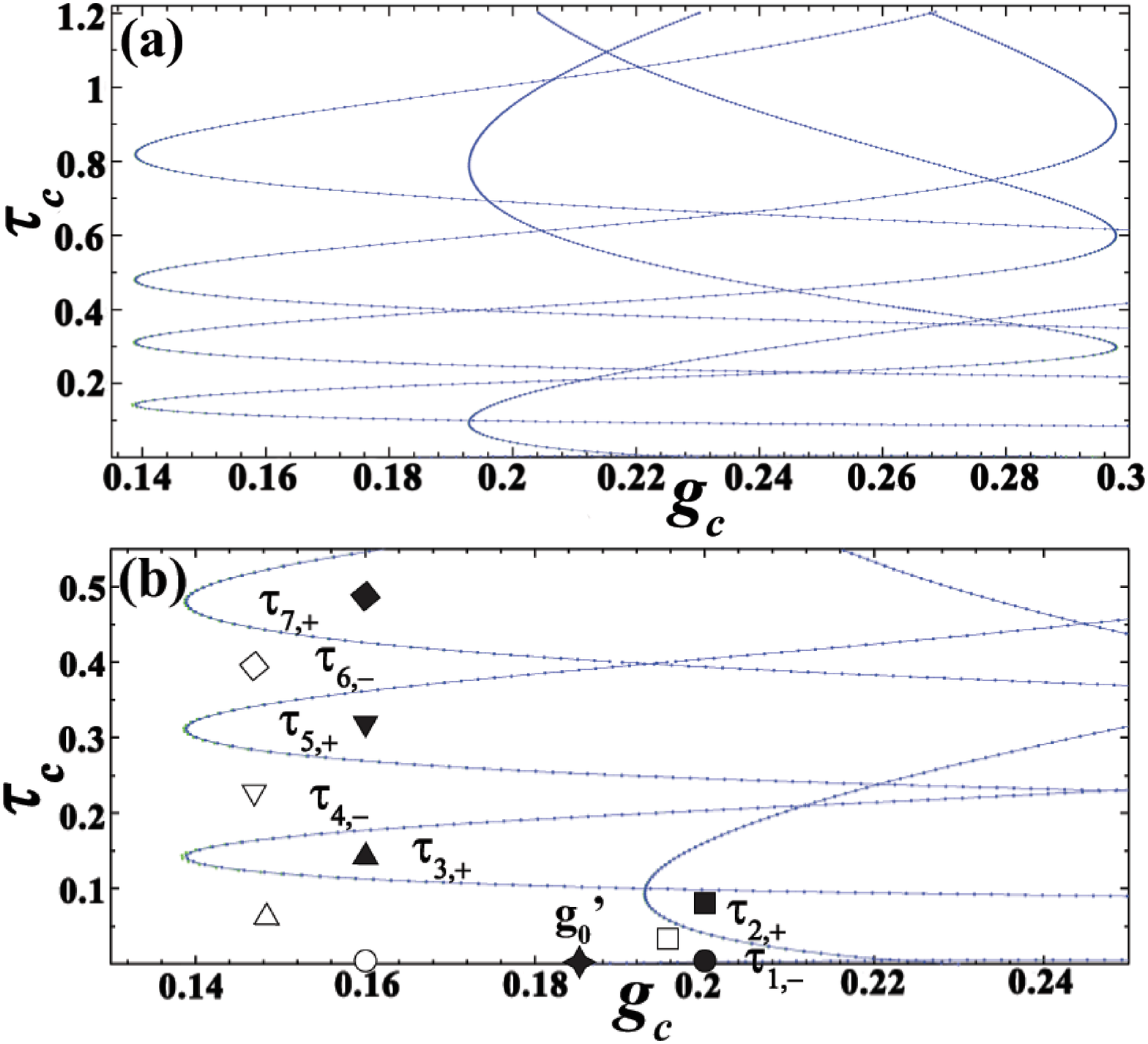}
%\end{minipage}
\caption{(color online) Bifurcation diagram in case of the symmetrical nonlinear cross-population coupling, as determined from the local bifurcation analysis of the approximate system made up of two identical interacting MF models. Unlike the conditions applied for Fig. \ref{Fig1}, the excitable units within each ensemble of the corresponding exact system are connected in the all-to-all fashion, with the couplings characterized by the non-negligible strength and time lag. (a) and (b) show the $\tau_c(g_c)$ Hopf bifurcation curves, whereby the close-up view in (b) is focused on the most relevant region of the parameter plane. The intrinsic parameters $D=0.0001, \tau_{in}=0.3$ and $g_{on}=0.1$ warrant that the isolated populations (single MF systems) exhibit the stochastically (asymptotically) stable stationary state.\label{Fig2}}
\end{figure*}

The main results in this Section concern the canonical setup involving two identical populations of interacting excitable neurons ($g_{in,1}=g_{in,2}=0.1$), whereby the cross-population couplings are taken to be symmetrical \cite{OPT10,MKB04}. The intrinsic ensemble parameters $D=0.0001,\tau_{in}=0.3$ warrant that the equilibrium is the only asymptotically (stochastically) stable state for the approximate (exact) model. Inspecting the appropriate bifurcation diagram in Fig. \ref{Fig2}(a), one readily realizes how, at variance with the previously discussed case, there is not one, but two scenarios for the destabilization of equilibrium. Which of the scenarios actually applies is contingent on the inter-population coupling strength $g_c$: if $g_c<g_0'$, viz. Fig. \ref{Fig2}(b), the equilibrium goes unstable via the direct supercritical Hopf bifurcation, while for $g_c>g_0'$, the onset of the collective mode rests with the direct subcritical Hopf bifurcation. In the latter instance, where $g_c$ notably outweighs $g_{in}$, an unstable limit cycle collapses at the fixed point making it unstable. Away from criticality, in the domain marked by the solid circle in Fig.\ref{Fig2}(b), the system's trajectory eventually gets drawn to a distant limit cycle attractor. Again, both the stable and the unstable limit cycle derive from the fold-cycle bifurcation, whereas the numerical simulations confirm that the unstable manifold of the equilibrium at $(g_c,\tau_c)=(g_0',0)$ supports the symmetrical oscillatory state. Below the curve $\tau_{1,-}$, which is barely distinguishable from the $g_c$-axes in Fig. \ref{Fig2}(b), one finds a narrow interval of coupling strengths $g_c\gtrsim g_0$ where the emanating branch of the unstable solutions apparently folds back. As a corollary, the system of coupled MF models is seen to exhibit a bistable regime, such that the equilibrium and the collective mode coexist. However, such bistability is difficult to observe in the dynamics of the full system for the sensitivity of the equilibrium to stochastic perturbation. Interestingly, the approximation for the critical coupling strength $g_0'$ is significantly improved when compared to the previous system configuration, this possibly owing to the influence of the inter-ensemble interactions that were excluded earlier on. Crossing into the domain $\tau_{1,-}<\tau_c<\tau_{2,+}$ represented by the open square in Fig. \ref{Fig2}(b), the MF system undergoes an inverse subcritical Hopf bifurcation, such that the fixed point loses an unstable plane. Looking in a more general picture, this region of parameter space is supposed to be bistable between the equilibrium and the large limit cycle born via the global bifurcation. In parallel, the unstable limit cycle from the Hopf bifurcation should act like a threshold for switching between the two solutions.  However, the stochastic component in the underlying dynamics prevents us from observing the bistable regime in the exact system. Above $\tau_{2,+}$, the equilibrium loses stability, giving way to the limit cycle as the sole attractor of the system's dynamics.

\begin{figure}[t]%2
%\begin{minipage}[t]{\columnwidth}
\centering
\includegraphics[scale=0.36]{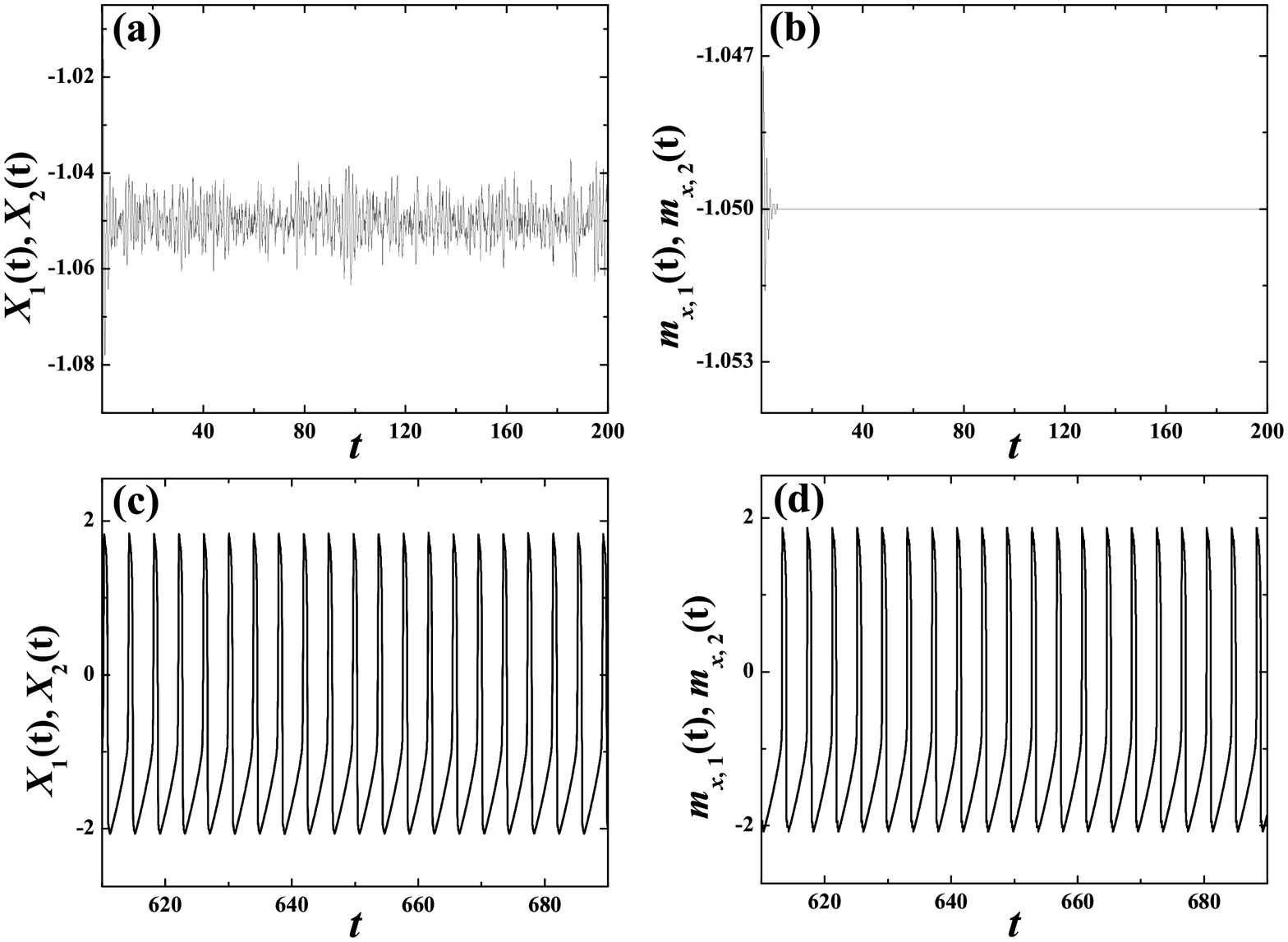}
%\end{minipage}
\caption{The exact (left column) and the approximate system (right column) are demonstrated to undergo the direct supercritical Hopf bifurcation when crossing the curve $\tau_{1,+}$ from Fig. \ref{Fig2}(b). (a) and (b) show that below the curve ($g_c=0.16,\tau_c=0.06$), the fixed point is stochastically stable for the exact, and asymptotically stable for the approximate system, respectively. The onset of oscillations above the curve ($g_c=0.16,\tau_c=0.14$)
is illustrated for the exact system in (c), and the approximate system in (d). The intrinsic population parameters are set to $D=0.0001, g_{in}=0.1$ and $\tau_{in}=0.3$.
\label{Fig3}}
\end{figure}

Next we turn to the sequence of bifurcations obtained for $g_c<g_0'$, which is a physically more plausible range since $g_c$ lies closer to $g_{in}$. Below $\tau_{3,+}$, the stationary state is stable for both the real and the approximate system, with the appropriate parameter domain highlighted by the open up-triangle in Fig. \ref{Fig2}(b). Crossing $\tau_{3,+}$ from below, the system undergoes the supercritical Hopf bifurcation, such that the equilibrium becomes unstable, and the emerging oscillations are symmetrical. An interesting point for the transition between the domains marked by the open and solid up-triangles in Fig. \ref{Fig2}(b) is that for the moderate coupling strength, under not too large a noise the time lag turns out to be a necessary ingredient should the equilibrium be destabilized. For the more comprehensive view, one again has to consider the effects of the interplay with the
global fold-cycle bifurcation, whereby a general remark is that everything stated on the direct supercritical Hopf bifurcation regarding the diagram in Fig. \ref{Fig1}(b) can carry over to this case. In brief, apart for the equilibrium, the system's phase space below $\tau_{3,+}$ also exhibits an unstable limit cycle enclosing the fixed point and a large stable limit cycle. Above the latter curve, the incipient limit cycle grows only until colliding with the unstable one, both being annihilated in an inverse fold-cycle bifurcation. Then, all the trajectories are eventually drawn to the large limit cycle, left as the sole attractor. As for the predictions of the approximate system, one stresses that there is an excellent agreement between the oscillating waveforms, in particular when comparing the anticipated frequency with the average one for the real system, viz. $T_{\blacktriangle,MF}=3.836$ vs. $T_{\blacktriangle,EX}=3.833$. This is illustrated in Fig. \ref{Fig3}, showing side-by-side the sequences from the time series $m_{x,i}(t)$ and $X_i(t)$ for $i=1,2$ below (top row) and above (bottom row) the curve $\tau_{3,+}$.

\begin{figure}[t]%2
%\begin{minipage}[t]{\columnwidth}
\centering
\includegraphics[scale=0.36]{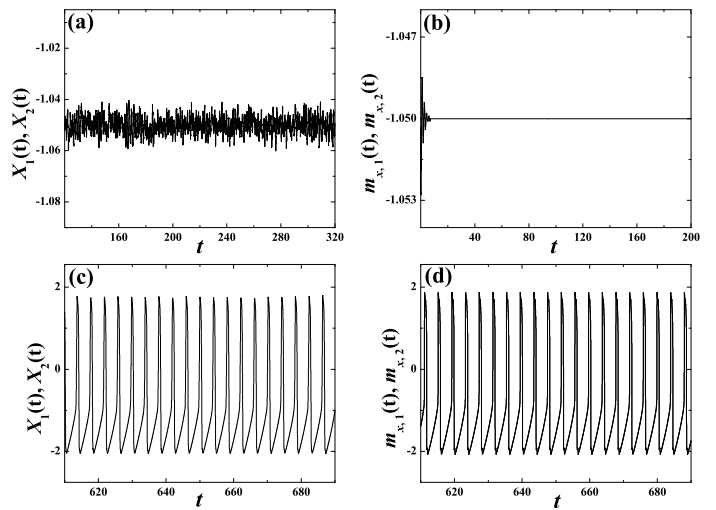}
%\end{minipage}
\caption{Illustration of how the same bistable regime, characterized by coexistence between the stationary and
the oscillatory state, is exhibited both by the exact (left column) and the approximate system (right column).
The top row indicates the corresponding stochastically and asymptotically stable fixed point, whereas the bottom row shows the two populations oscillating in-phase. The coupling strength and delay $(g_c,\tau_c)=(0.14,0.22)$ lie within the domain highlighted by the open down-triangle in Fig. \ref{Fig2}(b). The values for the intrinsic parameter subset are $D=0.0001, g_{in}=0.1$ and $\tau_{in}=0.3$.\label{Fig4}}
\end{figure}

Further enhancing $\tau_c$ to step into the domain highlighted by an open down-triangle in Fig. \ref{Fig2}(b), one encounters the bistable dynamics, such that the system, depending on the initial conditions, may display either the stationary or the oscillatory state. The area is bounded by $\tau_{4,-}$ from below and $\tau_{5,+}$ from above. The found bistability regime is the consequence of the inverse subcritical Hopf bifurcation, where the emanating unstable cycle effectively acts to stabilize the fixed point, allowing for it to coexist with the collective mode, the latter present due to the global bifurcation. The possibility of observing bistability in the exact system is likely facilitated by the unstable limit cycle, whose amplitude is sufficient to separate more clearly between the attraction basins of the oscillatory solution and the equilibrium in spite of the stochastic perturbations induced by the relatively small, but non-negligible noise. The bistable regime is illustrated in Fig. \ref{Fig4}, which demonstrates the coexistence of the stationary (top row) and oscillatory states (bottom row) for both the exact model and the MF approximation. Note that the change in oscillating frequency in the real system, associated with crossing $\tau_{4,-}$ from below, is well matched by the approximate system. Stepping into the domain $\tau_{5,+}<\tau_c<\tau_{6,-}$, marked by the solid down-triangle in Fig. \ref{Fig2}(b), the key change consists in the switch from the bistable to a monostable regime, the latter attributed the oscillatory state with the synchronization in-phase. The switch occurs as the system undergoes the direct supercritical Hopf bifurcation, which adds unstable directions, altering the stability of the fixed point. The change from the bistable to the monostable regime occurs in the same fashion for the MF and the exact system. Setting $\tau_c$ above $\tau_{6,-}$, see the domain represented by the open diamond in Fig. \ref{Fig2}(b), one finds the bistability regime reinstated. However, the transition is accompanied by the modulation of the oscillating frequency, the point well reflected by the approximate system, viz. $T_{\diamond,EX}=4.097$ against $T_{\diamond,MF}=4.119$. In general, the increase of coupling delay is biased toward reducing the oscillating frequency.

Note that the qualitatively similar sequence of bifurcations is verified to persist in a range of $g_{in}$ values, if $D$ and $\tau_{in}$ are set so to admit the stable stationary state as the sole attractor for the isolated populations. Nonetheless, in order for this framework to reflect accurately the behavior of the exact system, one should not consider too large noise amplitudes. The perturbation from larger $D$ may be envisioned as if leading to an effective broadening of the bifurcation curves for the real system, which renders the entire sequence smeared, and the underlying qualitative changes difficult to discern.

\begin{figure*}[t]%2
%\begin{minipage}[t]{\columnwidth}
\centering
\includegraphics[scale=0.4]{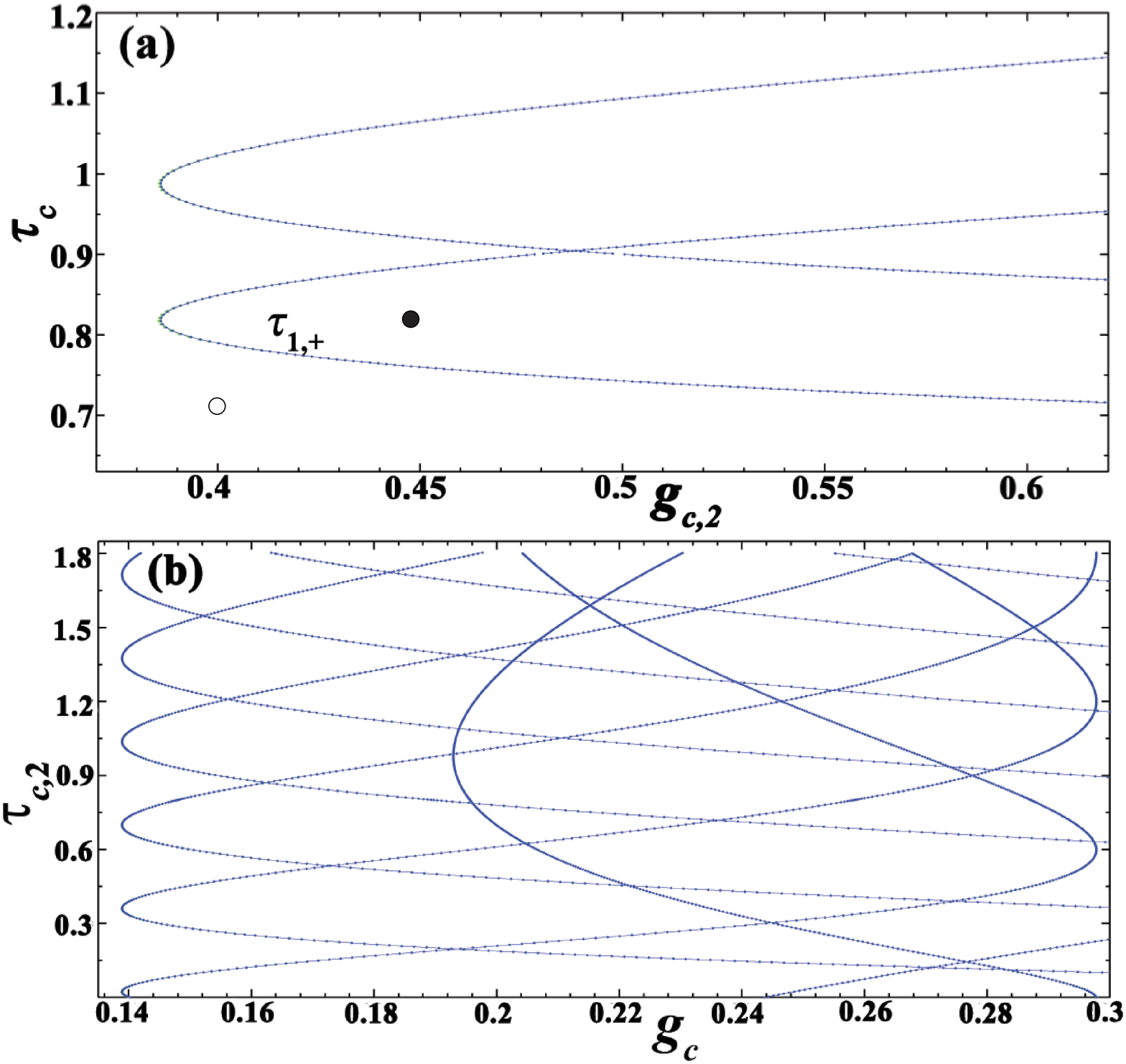}
%\end{minipage}
\caption{(color online) Results of the local bifurcation analysis carried out on the approximate system for the
two cases of asymmetrical cross-population couplings, are presented in the delay-strength parameter plane.
(a) refers to the setup with the disparate coupling strengths, holding $g_{c,1}=0.05$ and letting $g_{c,2}$ vary continuously. (b) is obtained for the uneven time lags, with $\tau_{c,1}=0.6$ fixed and $\tau_{c,2}$ allowed to change. The intrinsic parameters $D=0.0001,\tau_{in}=0.3,g_{in}=0.1$ are identical for both populations and warrant that the corresponding isolated system would exhibit the stationary state.\label{Fig5}}
\end{figure*}

{\it Asymmetrical coupling}

A question that naturally arises is whether and how is the physical picture so far modified by taking the asymmetrical, rather than the symmetrical cross-population coupling terms. We have examined two different scenarios: by one, the couplings in either direction retain a common time lag, but attain different strengths, whereas in the other, strengths are the same, but the transmission delays are disparate. In the former case, the coupling strength in one direction, say $g_{c,1}$ is kept fixed, while $g_{c,2}$ varies continuously. The bifurcation diagram in the $\tau_c$-$g_{c,2}$ plane is plotted in Fig. \ref{Fig5}(a), whereby the intrinsic population parameters are identical to those stated in the caption of Fig. \ref{Fig2}. One may immediately raise the issue of why is the bifurcation sequence profile much simpler compared to that in Fig. \ref{Fig2}(a). The possible reason lies in that for the cross-population couplings asymmetrical by strength, the system's behavior is predominantly influenced by the global bifurcation phenomena dependent on $g_{c,1}$ and $g_{c,2}$. Nonetheless, one cannot neglect some qualitative resemblance between the dynamics of the MF and the exact system. For instance, below $\tau_{1,+}$ in Fig. \ref{Fig5}(a), the equilibrium is stable for either system, but participates in the bistable regime. Along with the stationary state, one also finds an oscillatory state where the two populations are locked with a constant phase shift. This collective mode can only be attributed to the global bifurcation events. Crossing $\tau_{1,+}$ from below results in the creation of a limit cycle, leaving the equilibrium unstable. Both the real and the approximate model exhibit a single attractor supporting the phase-locked oscillations between the two populations, whereby the underlying frequencies are well matched, viz. $T_{\bullet,MF}=4.281$ against $T_{\bullet,EX}=4.302$. Notably, the oscillation waveforms above $\tau_{1,+}$ are more complex than those below, and bear the initial signatures of the quasiperiodic behavior. It has to be stressed that the qualitative resemblance between the dynamics of the exact and the approximate system heavily depends on how close is $g_{c,1}$ to $g_{in}$. In Fig. \ref{Fig5}(a), $g_{c,1}=0.05$ is comparably small to $g_{in}=0.1$. Should $g_{c,1}$ approach $g_{in}$ or exceed it, the effects of the global bifurcation phenomena become overwhelming, spoiling the predictions made by MF-based approximation.

We also briefly touch upon the setup where the cross-population couplings exhibit the disparate time lags, but attain the same coupling strength. Again, all the internal population parameters are equal to those linked to
Fig. \ref{Fig2}, whereas the notation on the asymmetrical coupling parameters is analogous to that used in the previous layout. The appropriate bifurcation diagram in the $\tau_{c,2}$-$g_c$ plane is displayed in Fig. \ref{Fig5}(b). Compared to Fig. \ref{Fig2}(a), we learn how the main difference between this case of asymmetrical couplings and the case with symmetrical interaction lies in the domain of small delays. In particular, the destabilization of equilibrium occurs solely via the supercritical Hopf bifurcation, whereas the scenario involving the subcritical Hopf bifurcation is absent. This picture seems to be independent on the relation between the fixed time lag $\tau_{c,1}$ and $\tau_{in}$.

Though it is not within the scope of the current study, one should still mention that the methods discussed can also be implemented for the scenarios where the two populations exhibit different types of kinetics, e.g. if one is made up of excitable, and the other of self-oscillating units. In this scenario, one effectively examines the interaction between the noise-induced and the noise-perturbed oscillations. The corresponding bifurcation diagram is not too distinct from the one in Fig. \ref{Fig5}(b), except that the pattern of bifurcation curves is less dense. The critical coupling strength analogous to $g_0$ is naturally smaller than the one for the interacting excitable populations. Nevertheless, this setup is distinguished from those considered earlier in that the unstable manifold of the equilibrium supports the onset of the collective mode with the phase-locked rather than the in-phase oscillations, such that the firing of the ensemble with self-oscillating neurons precipitates the firing of the ensemble containing the excitable neurons.

\section{Summary and discussion}\label{Summary}

In the present paper, we have pursued the analysis of the MF based approximation intended to accurately reflect the macroscopic behavior of two delay-coupled populations of stochastic excitable units in terms of the stability of the stationary state, the scenarios for the onset and the suppression of the collective mode, as well as the possibility of admitting bistable regimes, where the equilibrium and the oscillatory state are found to coexist. The described layout deserves attention, since it can be interpreted as the minimal model for the "network of networks", the configuration often brought into context of biological systems whose function relies on generation and adjustment between the multiple collective rhythms. The important ingredients of the exact system we consider include two types
of delayed interactions, whereby those within the ensembles are assumed to be linear, and the inter-ensemble ones, mediated by the appropriate global variables, are taken to be nonlinear. The corresponding approximate system is built by coupling the two MF models, derived to describe the activity of single populations. Such a framework follows the general idea that any ensemble of oscillating units exhibiting the collective mode can be treated as the macroscopic oscillator. The MF model integrates the cumulant approach with the Gaussian approximation, whereby the latter holds exactly if three conditions are satisfied. These include the thermodynamic limit $N\rightarrow\infty$ regarding the ensemble size, the negligible noise amplitude $D<<1$, as well as the negligible interaction between the units $g_{in}<<1$. However, it turns out that the approximate system is still able to predict with sufficient accuracy the behavior of large, but finite populations with non-negligible internal interactions, provided that the natural requirement for not too large a noise amplitude is met.

By stating the results in broad terms, the intention has been to stress their applicability to the class of systems made up of type II excitable units. Nonetheless, one recognizes that valuable motivation for the study comes from the field of neuroscience, which goes beyond the adopted model of local dynamics or the fashion in which the interactions are introduced. The methods for providing the reduced descriptions of the behavior of large neural assemblies are typically cast in the categories of the neural-mass and the MF models, whereby the former neglect, and the latter take into account the distribution of individual neuron states over the ensemble. In these terms, the model considered here interpolates between the two classes. Recall that we have introduced an additional approximation on the second order moments to translate the original MF system \eqref{eq7} into the form incorporated in \eqref{eq8}, with the latter preferred as it allows for the analytical tractability of the subsequent local bifurcation analysis.

An inference from such an analysis is that the approximate system can undergo direct and inverse supercritical or subcritical Hopf bifurcations, such that the direct (inverse) ones lead to the destabilization (stabilization) of the stationary state. The complex bifurcation sequence under variation of cross-population coupling strengths and delays is found to depend on the details of the system configuration, like the symmetrical or asymmetrical character of the bidirectional interaction between the ensembles. The main set of results refers to the symmetrical case, where it is demonstrated that the equilibrium may lose stability according to two different scenarios. One involves a direct supercritical Hopf bifurcation and can be achieved for instantaneous couplings solely by increasing $g_c$, whereas the other scenario unfolds via the direct subcritical Hopf bifurcation. The latter involves an interesting point that for strengths $g_c\simeq g_{in}$ one finds a time-lag threshold necessary to destabilize the equilibrium. Increasing $\tau_c$, there are parameter domains bounded from below (above) by the curves indicating subcritical (supercritical) bifurcations, where the stability of stationary state is regained. In many of such instances, the system is actually bistable, exhibiting coexistence between the equilibrium and the oscillatory state. This is a corollary of an interplay with the global fold-cycle bifurcation, as the large stable limit cycle born in this way remains unaffected by the local phenomena. Note that the global events may influence the system dynamics in several other instances. In particular, an unstable limit cycle created in a fold-cycle bifurcation may destabilize the fixed point in a direct subcritical Hopf bifurcation or may limit the growth of an incipient limit cycle following the direct supercritical Hopf bifurcation. By numerical simulation, we have verified that the parameter domains of stability or instability of equilibrium for the exact system are reproduced by the approximate one with high accuracy. In addition, it has been shown that the average oscillation frequency for the global variable of the exact system is well matched by that of the corresponding MF variable. In the exact system, the ability to observe the bistable regimes, where the unstable limit cycle act as a threshold between the equilibrium and the large cycle, is contingent on the noise amplitude. In general, the predictions of the approximate system are better if the deterministic component, governed by the coupling strengths and delays, prevails over the stochastic component in the dynamics of the exact system. An interesting study complementary to the present one would be to examine whether the MF based model may reproduce the forms of synchronization between the generated collective rhythms the way they are exhibited by the exact system. These could include the in-phase and antiphase synchronization or the phase-locked states, as well as their coexistence. The preliminary results implementing the H-function approach suggest that the approximate system may account for the stability of the synchronization regimes and provide indications on the possible multistability.

\begin{acknowledgments}
This work was supported in part by the Ministry of Education and
Science of the Republic of Serbia, under project Nos. $171017$ and
$171015$.
\end{acknowledgments}

\end{document}